\documentclass[12pt,reqno]{article}
\usepackage{amsmath}
\usepackage{amssymb}
\usepackage{hyperref}
\usepackage{color}

\hypersetup{colorlinks,linkcolor=blue,citecolor=red,urlcolor=blue,%
pdftitle={Membrane Instantons from Toda Field Theory},%
pdfauthor={Ulrich Theis (Riemann Center)}}

\allowdisplaybreaks
\numberwithin{equation}{section}

\renewcommand{\d}{\mathrm{d}}
\newcommand{\I}{\mathrm{i}}
\newcommand{\e}{\mathrm{e}}
\newcommand{\p}{\partial}
\newcommand{\mb}{\bar{m}}
\newcommand{\Ab}{\bar{A}}
\newcommand{\la}{\lambda}
\newcommand{\vq}{\vec{q}}
\newcommand{\vp}{\vec{p}}
\newcommand{\vu}{\vec{u}}
\newcommand{\vQ}{\vec{Q}}
\newcommand{\vR}{\vec{R}}
\newcommand{\sqr}{\sqrt{\smash[b]{r}\,}}
\newcommand{\tab}{\quad\,}
\newcommand{\cc}{\text{c.c.}}

\newcommand{\+}{\mskip1mu}
\renewcommand{\%}{\mskip-1mu}

\newcommand{\vc}[1]{\hat{#1}}
\newcommand{\frc}[2]{\frac{\raisebox{-1pt}{$#1$}}{#2}}
\newcommand{\arxiv}[1]{\href{http://arxiv.org/abs/#1}{arXiv:#1}}

\DeclareMathOperator{\re}{\mathrm{Re}}
\DeclareMathOperator{\im}{\mathrm{Im}}

\DeclareSymbolFont{AMSb}{U}{msb}{m}{n}
\DeclareMathSymbol{\fieldC}{\mathalpha}{AMSb}{"43}
\DeclareMathSymbol{\fieldN}{\mathalpha}{AMSb}{"4E}
\DeclareMathSymbol{\fieldP}{\mathalpha}{AMSb}{"50}
\DeclareMathSymbol{\fieldR}{\mathalpha}{AMSb}{"52}
\DeclareMathSymbol{\fieldZ}{\mathalpha}{AMSb}{"5A}

\definecolor{grey}{rgb}{.5,.5,.5}

\begin{document} 

\hfill ITP-UH-15/14
\vspace{4ex}

\begin{center}
 {\large\bfseries Membrane Instantons from Toda Field Theory} \\[3ex]
 {\large Ulrich Theis} \\[1ex]
 \slshape Institute of Theoretical Physics and Riemann Center for Geometry and Physics, \\ Leibniz Universit\"at Hannover, \\ Appelstr.~2, 30167 Hannover, Germany \\
 \href{mailto:ulrich.theis@riemann.uni-hannover.de}%
 {ulrich.theis@riemann.uni-hannover.de}
\end{center}

\vspace{5ex} \hrule \vspace{3ex}

Four-dimensional quaternion-K\"ahler metrics with an isometry are determined by solutions to the SU($\infty$) Toda equation. We set up a perturbation theory to construct approximate solutions to the latter which can be interpreted as membrane instanton corrections to the moduli space metric of the universal hypermultiplet. We compute one such solution exactly up to the five-instanton level, including all perturbative fluctuations about the instantons. The result shows a pattern that allows us to determine the asymptotic behaviour of \emph{all} higher instanton corrections within this solution. We find that the generating function for the leading terms of the latter is given by the Lambert $W$-function.

\vspace{3ex} \hrule \vspace{5ex}

\section{Introduction} 

Supersymmetry requires that the moduli space of hypermultiplets coupled to four-di\-men\-sional $N=2$ supergravity be quaternion-K\"ahler \cite{BW}. This applies in particular to the hypermultiplets in the low-energy effective action of Calabi--Yau threefold compactifications of Type II string theory. In the special case of a single hypermultiplet with a four-dimensional moduli space -- such as the universal hypermultiplet arising from the Type IIA theory on rigid Calabi--Yau threefolds -- the usual characterisation of quaternion-K\"ahler geometry in terms of its holonomy is empty and replaced by the condition that the metric be Einstein with non-zero scalar curvature and anti-self-dual Weyl tensor \cite{S}.

In this paper, we are particularly interested in such four-dimensional quaternion-K\"ahler metrics that admit an isometry, for reasons explained below. Przanowski \cite{P1} and Tod \cite{T1,T2} have found the canonical form of such metrics. In terms of local coordinates $(r,u,v,t)$ it reads
 \begin{equation}\label{Pmetric}
  G_4 = \frc{1}{r^2}\+ \big[ H\+ G_3 + H^{-1}\+ (\d t + \Theta)^2 \big]\ ,\quad G_3 = \d r^2 + \e^F (\d u^2 + \d v^2)\ .
 \end{equation}
The input is a function $F(r,u,v)$ satisfying the so-called SU($\infty$) Toda equation:
 \begin{equation} \label{Toda}
  \p_r^2\+\+ \e^F + \Delta\+ F = 0
 \end{equation}
with $\Delta=\p_u^2+\p_v^2$. From this one obtains the function $H(r,u,v)$ via the relation
 \begin{equation} \label{H}
  H = - \frc{3}{2\Lambda}\+ \big( 2 - r\+ \p_r F \big)\ ,
 \end{equation}
while the 1-form $\Theta(r,u,v)$ is a solution to
 \begin{align}
  \d \Theta & = *_3\+ (\d H + \omega\+ H)\ ,\quad \omega = \d r\, \p_r F \notag \\*[2pt]
  & = \p_r (\e^F\% H)\, \d u \wedge \d v + (\p_u H\, \d v - \p_v H\, \d u) \wedge \d r\ , \label{dtheta}
 \end{align}
the Hodge dual being taken with respect to the metric $G_3$. $\Lambda$ in \eqref{H} is the (target space) cosmological constant, i.e.\ $\text{Ric}_4= \Lambda\,G_4$; for the universal hypermultiplet $\Lambda=-3/2$.

Evidently, $G_4$ admits an isometry with Killing vector $\p_t$. From a physical point of view -- regarding $G_4$ as the target space metric of a non-linear sigma model -- this means that the scalar field $t$ can be dualised into a $(D-2)$-form $B$ subject to the (reducible) gauge transformation $B\rightarrow B+\d\lambda$ with $\lambda$ an arbitrary $(D-3)$-form. The target space metric of the remaining three-dimensional non-linear sigma model is then given by $r^{-2}\+H\+G_3$. As was observed in \cite{W}, $G_3$ is a representative metric of an Einstein--Weyl space with 1-form $2\omega$.

The classical universal hypermultiplet is described by the solution $\e^F=r$ to the Toda equation \eqref{Toda}. The resulting metric reads
 \begin{equation}
  G_4 = \frc{1}{r^2}\, \d r^2 + \frc{\mspace{2mu}1\mspace{2mu}}{r}\+ (\d u^2 + \d v^2) + \frc{1}{r^2}\+ (\d t + u\, \d v)^2\ ,
 \end{equation}
which corresponds to the non-compact unitary Wolf space $\mathrm{SU}(2,1)/ (\mathrm{SU}(2)\times\mathrm{U}(1))$. 
Its isometry group contains a Heisenberg subgroup generated by the Killing vectors $\p_t$, $\p_v$ and $\p_u-v\,\p_t$. Dualisation of $t$ yields the classical universal tensor multiplet, whose target space parameterized by $(r,u,v)$ is $\mathrm{SO}(3,1)/ \mathrm{SO}(3)$ -- i.e.\ Euclidean AdS$_3$ -- which we shall meet again below.

One may deform any solution $F(r,u,v)$ to \eqref{Toda} into another solution $F(r+c,u,v)$ with a real constant $c$. Applied to the classical universal hypermultiplet, one obtains $\e^F=r+c$, and the constant term captures the one-loop quantum correction to the moduli space metric 
with $c$ proportional to the Euler characteristic of the compactification manifold \cite{AMTV} (in fact, a non-renormalisation theorem ensures that there are no higher-order perturbative corrections that cannot be absorbed into field redefinitions \cite{RSV}). 
The Heisenberg isometry group is preserved by this correction.

The universal hypermultiplet also receives various non-perturbative quantum corrections, which arise from Euclidean branes wrapping cycles in the compactification manifold \cite{BBS} -- see e.g.\ \cite{BB,GS,TV,K1} for early approaches of computing them. These will in general break all isometries of the target space and the Przanowski--Tod metric is no longer applicable. This situation was investigated in \cite{ASV}, where membrane and fivebrane corrections at the one-instanton level were derived within a more general framework also developed by Przanowski \cite{P2} (see also \cite{APV,H}). However, if one considers corrections only from membrane instantons, the axion shift symmetry generated by $\p_t$ survives and there should be a solution to the Toda equation that describes the non-perturbatively corrected metric. This is the motivation for our present investigation. While less ambitious than the work of \cite{ASV}, the preserved isometry allows us to go beyond the one-instanton level -- far beyond in fact in the presence of a further isometry. In \cite{DSTV} the SU($\infty$) Toda equation had already been used to determine possible membrane instanton corrections, but these were limited to just a few of the leading terms in the one- and two-instanton sector.

In fact, after it was shown in \cite{RLRSTV} how to use string dualities to determine some of the non-perturbative corrections to the hypermultiplet moduli space exactly, the latter have by now all been worked out to a large extent in a series of papers \cite{APSV,A1,APP,AB} (see \cite{A2} for a review and further references), but the twistor construction employed there makes the derivation of an explicit moduli space metric cumbersome, if not impractical -- to our knowledge, no metric has actually been computed yet from the results in the above references. This is why we still regard the construction of suitable solutions to the Toda equation as a worthwile endeavour. Moreover, a few years ago it has been realised that four-dimensional anti-self-dual Einstein metrics determined by the Toda equation also occur in a different context as Einstein-Maxwell instantons in Euclidean gauged $N=2$ supergravity \cite{DGST} and provide a good testing ground for the gauge/gravity correspondence \cite{FLMS}. (The earliest applications to Euclidean gravity appear in \cite{B1,P}.)

Various solutions to the SU($\infty$) Toda equation are known (see e.g.~\cite{W,LB,BS,B2,CT,LLM}), but none of these appears to be relevant to the problem at hand. We cannot exclude the possibility of a suitable coordinate transformation or resummation turning an already known solution into a form that does correspond to instanton corrections in a manifest way, though we expect that finding such a transformation -- if it exists -- is no less difficult than deriving an instanton solution from scratch. We have chosen the latter approach.
\bigskip

The outline of the paper is as follows: In section \ref{sec:instdef} we consider an Ansatz for an additive deformation of the classical solution $\e^F=r$ to the Toda equation and set up a perturbation theory to solve the latter. Crucial to the procedure is a reformulation of the Toda equation in the background of this classical solution into an inhomogeneous Laplace equation in Euclidean AdS$_3$. Solving this to first order in the deformation, we find that membrane instanton-like functions appear automatically. At second order in what we interpret as the instanton number, certain obstructions arise. We determine a consistent deformation to this order, i.e.\ an exact two-instanton deformation of the classical solution.

In section \ref{sec:def2isom} we restrict ourselves to a special case of the previous deformation which admits a second isometry, allowing us to continue the iterative procedure up to the five-instanton level (and beyond if desired). While we have been unable to solve the Toda equation to all orders and thereby derive an exact solution, we present an empirical formula for higher order deformations which correctly reproduces the leading terms in \emph{all} instanton/anti-instanton sectors, which we determine rigorously in the appendices. Remarkably, we find that rational contributions to these leading terms always sum up to integers. We conclude by showing that the generating function for the dominant term in each instanton sector of this particular deformation is given by the Lambert $W$-function.

It should be noted that for four-dimensional anti-self-dual Einstein metrics with two commuting isometries the Przanowski--Tod framework is not the optimal one -- Calderbank and Pedersen have found the most general form of such metrics in \cite{CP}, which are determined by a function that satisfies a \emph{linear} partial differential equation much simpler than the Toda equation. It is nevertheless beneficial to use the Przanowski--Tod form, since it allows for a generalisation of metrics with two isometries by breaking one of them, of which the aforementioned two-instanton deformation in section \ref{sec:instdef} provides an example.

\section{Instanton deformations} 
\label{sec:instdef}

In Calabi--Yau compactifications of Type IIA string theory, the real part of a membrane instanton action is of the form $2\+|\vQ|\,g_s^{-1}$, where $g_s$ is the string coupling constant and $\vQ$ is a two-dimensional charge vector whose integral components determine how the membrane wraps around 3-cycles. As explained in \cite{DSTV}, only charge vectors with one component vanishing should contribute. $g_s$ is set by the vacuum expectation value of the dilaton in the universal hypermultiplet, which is described by the $r$ coordinate -- the precise relation being $\sqr=g_s^{-1}$. We shall therefore be looking for solutions to the Toda equation that in the small-coupling -- i.e.\ large $r$ -- limit fall off exponentially like $\e^{-2n\+\sqr}$ in the $n$-instanton sector with $n\in\fieldN$.

In order to solve the Toda equation iteratively, it is advantageous to consider $\e^F$ as the function to be determined rather than $F$ itself. Let us first separate the classical solution from the non-per\-tur\-ba\-tive corrections (as mentioned above, the one-loop correction can be incorporated by a constant shift $r\rightarrow r+c$):
 \begin{equation}
  \e^{F(r,\vu\+)} = r - f(r,\vu\+)
 \end{equation}
with $\vu=(u,v)$. A few manipulations then turn the Toda equation \eqref{Toda} into
 \begin{equation} \label{Toda2}
  (r\+ \p_r^2 + \Delta)\+ f = \Delta \big[ f + r \ln(1 - f/r) \big] = -r\, \Delta \sum_{k=2}^\infty\, \frc{1}{k}\, (f/r)^k\ .
 \end{equation}
In this form the equation resembles an inhomogeneous massless Klein-Gordon equation in Euclidean AdS$_3$. Indeed, in terms of the variable $x=2\sqr$, the operator
 \begin{equation} \label{Box}
  \Box = 4\+ r\+ (r\+ \p_r^2 + \Delta) = x^2 \p_x^2 - x\+ \p_x + x^2 \Delta
 \end{equation}
coincides with the (negative) Laplacian on the upper half-space $H=\fieldR^3|_{x>0}$ equipped with the Poincar\'e metric $ds^2=x^{-2}\+ (\d x^2+\d\vu\+\+{}^2)$. Up to a numerical factor the latter is just the target space metric $r^{-2}\+h\,G_3$ of the classical universal tensor multiplet.\footnote{In \cite{LB} LeBrun already made use of the Laplacian on hyperbolic space in the context of the SU($\infty$) Toda equation, but only for the \emph{linearised} equation $\p_r^2(\e^F\%H)+\Delta H=0$, which is the integrability condition for the relation \eqref{dtheta} determining $\Theta$.}

Let us now decompose $f$ into functions $f_n$ supposed to describe the $n$-instanton sector, in the sense that we are looking for solutions in which for large $x$ the $f_n$ fall off exponentially like $\e^{-nx}$ as explained above:
 \begin{equation} \label{fn}
  f(r,\vu\+) = \sum_{n=1}^\infty\+ f_n(r,\vu\+)\ .
 \end{equation}
Expanding \eqref{Toda2} in the instanton number $n$ and equating terms of the same order using Fa\`a di~Bruno's formula, we find
 \begin{equation} \label{Boxfn}
  \Box\+ f_n = -4\+ r^2\+ \Delta\! \sum_{\sum_{\ell=1}^{n-1} \ell\, k_\ell = n}\! \frc{(k_1 + \dots + k_{n-1} - 1)!}{k_1! \dots k_{n-1}!}\, \frc{f_1^{k_1} \dots f_{n-1}^{k_{n-1}}}{r^{k_1 + \dots + k_{n-1}}}\ .
 \end{equation}
The total number of terms in this equation, including the one on the left-hand side, is given by the number of partitions of $n$. The summation index $k_\ell$ counts the multiplicity of the number $\ell<n$ in each partition. Explicitly, we have up to fifth order (we give a solution to this order below)
 \begin{subequations}
 \begin{align}
  \Box\+ f_1 & = 0 \label{T1} \\[2pt]
  \Box\+ f_2 & = -\frc{4}{2}\, \Delta\+ f_1^2 \label{T2} \\[2pt]
  \Box\+ f_3 & = -\frc{4}{3\+r}\, \Delta\+ \big( f_1^3 + 3\+r f_1 f_2 \big) \\[2pt]
  \Box\+ f_4 & = -\frc{4}{4\+r^2}\, \Delta\+ \big( f_1^4 + 4\+r f_1^2 f_2^{} + 4\+r^2 f_1 f_3 + 2\+ r^2 f_2^2 \big) \\[2pt]
  \Box\+ f_5 & = -\frc{4}{5\+r^3}\, \Delta\+ \big( f_1^5 + 5\+r f_1^3 f_2^{} + 5\+r^2 f_1^2 f_3^{} + 5\+r^2 f_1^{} f_2^2 + 5\+r^3 f_2 f_3 + 5\+r^3 f_1 f_4 \big)\ . \label{T5}
 \end{align}
 \end{subequations}

In order to solve \eqref{T1}, we consider the Fourier transform with respect to the $\vu$ variables
 \begin{equation}
  f_1(x,\vu\+) = \int\! d^2\%q\ \e^{\+\I\+ \vq\+ \cdot\+ \vu}\+ \tilde{f}_1(x, \vq\+)\ .
 \end{equation}
$\tilde{f}_1$ then has to satisfy
 \begin{equation}
  x^2 \tilde{f}_1'' - x \tilde{f}_1' - q^2 x^2 \tilde{f}_1 = 0
 \end{equation}
with $q=|\vq\+|$, which for $q\neq 0$ is solved by the modified Bessel functions $x\+K_1(qx)$ and $x\+I_1(qx)$ (the case $q=0$ gives the perturbative solution linear in $r$). For large $x=2\sqr$ -- which as we recall corresponds to a small string coupling constant -- the second solution diverges exponentially, so it should be discarded. On the other hand, for large argument the asymptotic behaviour of the functions $K_\nu(z)$ is given by
 \begin{equation} \label{Kasympt}
  K_\nu(z) \sim \e^{-z} \sqrt{\pi/2z}\, \Big( 1 + \sum_{k=1}^\infty \frac{a_k(\nu)}{z^k} \Big)\ ,\quad a_k(\nu) = \frc{1}{8^k\+ k!}\, \prod_{\ell=1}^k \big( 4\nu^2 - (2\ell - 1)^2 \big)\ ,
 \end{equation}
the leading term being just what we expect for an exponentially suppressed instanton. In the field or string theory context the asymptotic power series would be interpreted as perturbative quantum corrections due to fluctuations about the instanton background. The modified Bessel functions $K_\nu(z)$ are ubiquitous in stringy instanton calculations as they arise in the Poisson resummation of Eisenstein series -- see e.g.~\cite{OV,GG,GV,K1}.

Thus, the general solution to \eqref{T1} with the desired asymptotics is given by
 \begin{equation} \label{f1}
  f_1(x,\vu\+) = x\! \int\! d^2\%q\ \e^{\+\I\+ \vq\+ \cdot\+ \vu}\+ K_1(qx)\, \varrho_1(\vq\+)\ .
 \end{equation}
Here $\varrho_1(\vq\+)$ is an as yet arbitrary function (more generally a distribution, see below) subject to the reality condition $\varrho_1^*(\vq\+)= \varrho_1(-\vq\+)$. From the asymptotics we read off the instanton action $S=2q\+\sqr-\I\+\vq\cdot\%\vu$. Due to the $\vq\+$-integral, $f_1$ contains ``instantons'' of all -- even continuous -- orders, which calls into question the above decomposition of the Toda equation. Some constraint has to be imposed on the function $\varrho_1$ to fix the instanton order. In fact, the equation \eqref{T2} at the next instanton level will restrict $\varrho_1$, but not sufficiently. It was not to be expected that the Toda equation alone would impose instanton charge quantisation, this has to be put in by hand.

The homogeneous solution for each $f_n$ is of the same form \eqref{f1}. Particular in\-homogeneous solutions to the higher order equations can in principle be found by means of the Green's function of the operator \eqref{Box}, which satisfies
 \begin{equation}
  \Box\, G(\vc{x},\vc{x}') = - x^3\+ \delta(\vc{x} - \vc{x}')\ .
 \end{equation}
Here we use the notation $\vc{x}=(x,\vu\+)$. Explicitly, it reads (see e.g.\ section 6.3 in \cite{DHF})
 \begin{equation}
  G(\vc{x},\vc{x}') = \frc{\xi^2}{8\pi}\, {}_2F_1(1\+, \tfrac{3}{2}\+; 2\+; \xi^2) = \frc{\xi^2}{8\pi}\, \sum_{k=0}^\infty\, \frc{(2k+1)!!}{2^k\+ (k+1)!}\ \xi^{2k} = \frc{\xi^2}{4\pi \big( 1 - \xi^2 + \sqrt{1 - \xi^2}\+ \big)}\ ,
 \end{equation}
where
 \begin{equation}
  \xi(\vc{x},\vc{x}'\+) = \frc{2 x x'}{x^2 + x'{}^2 + (\vu - \vu{\+}')^2}\ .
 \end{equation}
The general instanton-like solution to the inhomogeneous equation
 \begin{equation}
  \Box\+ f_n(\vc{x}\+) = J_n(\vc{x}\+)
 \end{equation}
is then given by
 \begin{equation}
  f_n(\vc{x}) = x\! \int\! d^2\%q\ \e^{\+\I\+ \vq\+ \cdot\+ \vu}\+ K_1(qx)\, \varrho_n(\vq\+) - \int_H \frc{d^3\%y}{y^3}\ G(\vc{x}, \vc{y})\+ J_n(\vc{y})\ .
 \end{equation}

We can actually find some particular solutions to \eqref{T2} without the help of the Green's function. Inserting $f_1$, the equation turns into
 \begin{equation} \label{Boxf2}
  \Box\+ f_2 = -2\+ \Delta f_1^2 = 2\! \int\! d^2\%p \int\! d^2\%q\ (\vp + \vq\+)^2\, \e^{\+\I (\vp + \vq\+) \cdot \vu}\+ \varrho_1(\vp\+)\, \varrho_1(\vq\+)\, x^2 K_1(px)\+ K_1(qx)\ .
 \end{equation}
Let us assume that a solution to the inhomogeneous equation is of the form\footnote{A comparison with \cite{AMNP} suggests that this Ansatz might be too simple to describe the universal hypermultiplet \cite{A3}.}
 \begin{equation} \label{f2}
  f_2(x,\vu\+) = \int\! d^2\%p \int\! d^2\%q\ \e^{\+\I (\vp + \vq\+) \cdot \vu}\+ \varrho_1(\vp\+)\, \varrho_1(\vq\+)\, \hat{f}_2(x, \vp, \vq\+)\ ,
 \end{equation}
where $\hat{f}_2$ then has to satisfy
 \begin{equation} \label{Boxhf2}
  x^2 \hat{f}_2'' - x \hat{f}_2' - (\vp + \vq\+)^2\+ x^2 \hat{f}_2 = 2 (\vp + \vq\+)^2\+ x^2 K_1(px)\+ K_1(qx)\ .
 \end{equation}
The derivative identities
 \begin{equation}
  \p_z\+ K_0(z) = -K_1(z)\ ,\quad \p_z\+ (z\+ K_1(z)) = -z\+ K_0(z)
 \end{equation}
suggest to make an Ansatz
 \begin{align}
  \hat{f}_2(x, \vp, \vq\+) & = a(\vp, \vq\+)\, x\+ K_0(px)\+ K_1(qx) + b(\vp, \vq\+)\, x\+ K_0(qx)\+ K_1(px) \notag \\*[2pt]
  & \tab + c(\vp, \vq\+)\, x^2\+ K_0(px)\+ K_0(qx) + d(\vp, \vq\+)\, x^2\+ K_1(px)\+ K_1(qx)
 \end{align}
with $b(\vp,\vq\+)=a(\vq,\vp\+)$. We then find the following conditions arising from the different products of Bessel functions appearing in \eqref{Boxhf2}:
 \begin{subequations}
 \begin{align}
  K_1(px)\+ K_1(qx) :\quad & p\+ a + q\+ b = (\vp + \vq\+)^2 \label{KK1} \\[2pt]
  x\+ K_0(px)\+ K_1(qx) :\quad & pq\+ b - q\+ c = \vp \cdot \vq\ a \label{KK2} \\[2pt]
  x\+ K_1(px)\+ K_0(qx) :\quad & pq\+ a - p\+ c = \vp \cdot \vq\ b \label{KK3} \\[2pt]
  x^2\+ K_0(px)\+ K_1(qx) :\quad & pq\+ d = \vp \cdot \vq\ c \label{KK4} \\[2pt]
  x^2\+ K_1(px)\+ K_0(qx) :\quad & pq\+ c = \vp \cdot \vq\ d\ . \label{KK5}
 \end{align}
 \end{subequations}

Equations \eqref{KK1} to \eqref{KK4} are solved by
 \begin{align}
  a(\vp, \vq\+) & = p^{-1}\, \vp\+ \cdot\% (\vp + \vq\+)\ ,\quad c(\vp, \vq\+) = (pq)^{-1}\+ \big( p^2\+ q^2 - (\vp \cdot \vq\+)^2 \big) \notag \\[2pt]
  b(\vp, \vq\+) & = q^{-1}\, \vq\+ \cdot\% (\vp + \vq\+)\ ,\quad d(\vp, \vq\+) = (pq)^{-1}\+ (\vp \cdot \vq\+)\, c(\vp, \vq\+)\ .
 \end{align}
A possible addition of $e(\vp\+)\,\delta(\vp+\vq\+)$ to $a$ with an arbitrary function $e(\vp\+)=-e(-\vp\+)$ would drop out of $\hat{f}_2$. Upon inserting $c$ and $d$ into \eqref{KK5} and restoring the integrals, we find that \eqref{f2} solves \eqref{Boxf2} provided the following constraint is satisfied:
 \begin{equation} \label{cons}
  \int\! d^2\%p \int\! d^2\%q\ \varrho_1(\vp\+)\, \varrho_1(\vq\+)\, \e^{\+\I (\vp + \vq\+) \cdot \vu}\, \big( p^2\+ q^2 - (\vp \cdot \vq\+)^2 \big)^2\, (p\+q)^{-2}\+ K_1(px)\+ K_0(qx) = 0\ .
 \end{equation}

This restricts the possible choices of $\varrho_1(\vq\+)$. Until now we have tacitly assumed it to be an ordinary function. This does not appear to be compatible with the above constraint. The only way $\varrho_1(\vq\+)$ can relate $\vq\+$ to $\vp$, which are independent integration variables, is to fix the direction of $\vq$ and hence of $\vp\,$. This is achieved by a distribution of the form
 \begin{equation}
  \varrho_1(\vq\+) = \int\! d\la\ D_1(\la)\, \delta\big( \vq - \la \vQ\+ \big)\ ,
 \end{equation}
where $\vQ\in\fieldR^2$ is some constant \emph{unit} charge vector and $D_1(\la)=D_1^*(-\la)$ may contain derivatives with respect to $\vq\+$. It is not clear to us how the same condition arises from the Green's function approach. We shall consider $D_1(\la) \propto \delta(\la\pm 1)$ only, which allows us to speak of \emph{one} instanton.

If $D_1(\la)$ with $|\la|=1$ contains no derivatives, i.e.
 \begin{equation} \label{g1}
  \varrho_1(\vq\+) = \frc{1}{2}\+ A\, \delta\big( \vq - \vQ\+ \big) + \frc{1}{2}\+ \Ab\, \delta\big( \vq + \vQ\+ \big)
 \end{equation}
with $A\in\fieldC$ constant\footnote{The prefactor $1/2$ avoids numerous powers of 2 in the next section.}, the constraint \eqref{cons} is satisfied by virtue of $\vq$ and $\vp$ being enforced to be parallel. Observe that $c(\vp,\vq\+)$ and $d(\vp,\vq\+)$ then vanish. At least up to the first instanton order the resulting metric will possess a second isometry, commuting with $\p_t$, since $u$ and $v$ then appear only in the combination $\vQ\%\cdot\%\vu$ (provided the 1-form $\Theta$ is suitably chosen). We shall study this case in detail in the next section.

In order to break the isometry in the $(u,v)$-sector, $\varrho_1(\vq\+)$ has to contain derivatives of delta functions. Consider
 \begin{equation}
  \varrho_1(\vq\+) = \frc{1}{2}\+ \big( A + \I\+ \vR \cdot\% \nabla_{\!\vq}\+ \big)\, \delta\big( \vq - \vQ\+ \big) + \frc{1}{2}\+ \big( \Ab + \I\+ \vR^*\% \cdot\% \nabla_{\!\vq}\+ \big)\, \delta\big( \vq + \vQ\+ \big)
 \end{equation}
with another constant vector $\vR\in\fieldC^2$. Inserted into the constraint \eqref{cons}, the derivatives act on everything to the right of $\varrho_1(\vp\+)\, \varrho_1(\vq\+)$. Unless both factors of the square $\big( p^2\+ q^2 - (\vp \cdot \vq\+)^2 \big)^2$ are being differentiated, at least one of them vanishes due to the condition $\vq=\pm\vp\+$ imposed by the delta functions. It remains to evaluate the terms of the form
 \begin{align}
  & \nabla_{\!\vp}\, \big( p^2\+ q^2 - (\vp \cdot \vq\+)^2 \big) \otimes \nabla_{\!\vq}\, \big( p^2\+ q^2 - (\vp \cdot \vq\+)^2 \big) \notag \\[2pt]
  & = 4\+ \big( q^2\+ \vp - (\vp \cdot \vq\+)\, \vq\+ \big) \otimes \big( p^2\+ \vq - (\vp \cdot \vq\+)\, \vp\+ \big)\ .
 \end{align}
These too vanish once the integrations are performed and hence the constraint is satisfied. 
$\varrho_1(\vq\+)$ then yields the following solution to \eqref{T1}:
 \begin{equation}
  f_1(x,\vu\+) = \frc{x}{2}\+ \big[ \big( A + \vR\% \cdot\% \vu\+ \big)\+ K_1(x) + \I\+ \vR \cdot\% \vQ\, \big( K_1(x) + x\+ K_0(x) \big) \big]\+ \e^{\+\I \vQ \cdot \vu} + \cc\ .
 \end{equation}
Clearly, if $\vR$ is not aligned with $\vQ$, the first term in $f_1$ will break the remaining isometry in the $(u,v)$-sector. In the special case $\vR\cdot\vQ=0$, $f_1$ reduces to the function in equation (C.5) in \cite{ASV}. Note that this condition allows for the discrete symmetry transformation $\vu\rightarrow \vu+2\pi\+m\+\vQ$ with $m\in\fieldZ$ (accompanied by a transformation of $t$ whose form depends on the choice of $\Theta$).

$f_2$ can be computed straightforwardly from the above formulae. $c(\vp,\vq\+)$ and $d(\vp,\vq\+)$ now contribute since they get differentiated twice before the condition $\vq=\pm\vp\+$ is imposed. Let us give the result only for the case $\vR\cdot\vQ=0$, which is already rather complicated:
 \begin{align}
  f_2(x,\vu\+) & = \frc{x}{2}\+ \big[ \big( 2\+ (A + \vR\% \cdot\% \vu\+)^2 - R^2 \big)\+ K_1(x)\+ K_0(x) + x\+ R^2\+ \big( K_0(x)^2 + K_1(x)^2 \big) \big]\+ \e^{\+2\I \vQ \cdot \vu} \notag \\*
  & \tab - \frc{x}{2}\, |R|^2\+ \big[ K_1(x)\+ K_0(x) + x\+ \big( K_0(x)^2 - K_1(x)^2 \big) \big] + \cc\ . \label{f2R}
 \end{align}
Each term is bilinear in Bessel functions and behaves asymptotically like $\e^{-2x}$ for large $x$, in agreement with our identification of $f_2$ as a 2-instanton deformation. We comment on the absence of a phase factor in the second line in the next section. 

The metric $G_4$ can now be determined from the formulae in the introduction, the ingredients being the function $H$ and the 1-form $\Theta$ apart from $F$. We again restrict ourselves to the case $\vR\cdot\vQ=0$ and consider only the 1-instanton sector. As mentioned above, the perturbative corrections can be incorporated by means of a constant shift $r\rightarrow r+c$ in $F$:
 \begin{equation}
  \e^{F(r,\vu\+)} = r + c - \sqrt{r+c\,}\, K_1(2\sqrt{r+c\,})\+ \big[ (A + \vR\% \cdot\% \vu\+)\, \e^{\+\I \vQ \cdot \vu} + \cc \big] + \dots\ .
 \end{equation}
Here the ellipsis denotes higher instanton contributions. We then obtain from \eqref{H} (with $\Lambda=-3/2$)
 \begin{equation}
  H(r,\vu\+) = 1 + \frc{c}{r+c} - \frc{r}{r+c}\, K_2(2\sqrt{r+c\,})\+ \big[ (A + \vR\% \cdot\% \vu\+)\, \e^{\+\I \vQ \cdot \vu} + \cc \big] + \dots\ ,
 \end{equation}
where we have used the identity
 \begin{equation} \label{K2}
  \frc{2}{z}\+ K_1(z) + K_0(z) = K_2(z)\ .
 \end{equation}
Finally, up to an exact form the solution to \eqref{dtheta} is given by
 \begin{align}
  \Theta(r,\vu\+) = \frc{1}{2}\, \vu \times \d\vu\+ - \big[ & K_0 (2\sqrt{r+c\,}) + \frc{r}{\sqrt{r+c\,}}\, K_1(2\sqrt{r+c\,}) \big]\, \cdot \notag \\
  \cdot\, \big[ & \big( \vR + \I\+ (A + \vR\% \cdot\% \vu\+)\, \vQ\+ \big)\, \e^{\+\I \vQ \cdot \vu} + \cc \big]\% \times \d\vu\+ + \dots\ .
 \end{align}

\section{A deformation with two isometries} 
\label{sec:def2isom}

We now have a closer look at the deformation following from $\varrho_1(\vq\+)$ in \eqref{g1}. This is simple enough to go beyond the 2-instanton level. In terms of the variable $w=\vQ\%\cdot\%\vu$ we then have
 \begin{equation}
  f_1(x,w) = \frc{x}{2}\, A_1\, K_1(x)\, \e^{\+\I w} + \cc\ ,
 \end{equation}
where we write $A_1$ instead of $A$ for the time being. To preserve the second isometry at higher orders, we make the same simplifying Ansatz for every homogeneous solution $f_n$, only with instanton charge $n$. At each level we then obtain one new complex constant $A_n$. This leads to the following 2-instanton deformation:
 \begin{equation}
  f_2(x,w) = \frc{x}{2}\+ \big[ A_2\, K_1(2x) + 2\+ A_1^2\, K_1(x)\+ K_0(x) \big]\+ \e^{\+2\I w} + \cc\ .
 \end{equation}

We can go on and solve the equations of higher order as well without encountering any more obstructions, but the number of terms grows quickly. At the 3-instanton level, some new structure appears:
 \begin{align}
  f_3(x,w) & = \frc{x}{2}\+ \big[ A_3\, K_1(3x) + 3\+ A_1 A_2\+ \big( K_1(x)\+ K_0(2x) + K_0(x)\+ K_1(2x) \big) \notag \\*
  & \mspace{44mu} + \tfrac{3}{2}\+ A_1^3\, K_1(x) \big( K_1(x)^2 + 3\+ K_0(x)^2\+ \big) \big]\+ \e^{\+3\I w} \notag \\[4pt]
  & \tab + \frc{x}{2}\+ \big[ \Ab_1 A_2\+ \big( K_1(x)\+ K_0(2x) - K_0(x)\+ K_1(2x) \big) \notag \\*
  & \mspace{63mu} + \tfrac{1}{2}\+ A_1^2 \Ab_1\, K_1(x) \big( K_1(x)^2 - K_0(x)^2\+ \big) \big]\+ \e^{\+\I w} + \cc\ .
 \end{align}
Note that while all terms go like $\e^{-3x}$ for large $x$, there are two different phase factors similar to what we have already observed in \eqref{f2R}. These phase factors allow us to distinguish instanton and anti-instanton contributions, counted by pairs of non-negative integers $(m,\mb)$. $n$-instanton terms with $n=m+\mb$ carry net charge $q=m-\mb$. $f_3$ contains $(3,0)$- and $(2,1)$-instanton contributions as well as their complex conjugates with $m$ and $\mb$ interchanged. In the above expressions we observe that no terms with $m=\mb$ occur\footnote{While such terms do arise on the right-hand side of the equation \eqref{Boxfn}, they are independent of $w$ and hence annihilated by $\Delta$.} -- the instantons at hand all carry non-zero net charge. By contrast, the 2-instanton deformation \eqref{f2R} in the previous section also contains a $(1,1)$-instanton contribution.

\begin{table}[t]
 \begin{equation*}
  \begin{array}{c|ccccccccccc}
  n\setminus q & 5 & 4 & 3 & 2 & 1 & 0 & -1 & -2 & -3 & -4 & -5 \\[.5ex]
  \hline \rule[6pt]{0cm}{2ex}
  0 & & & & & & (0,0) & & & & & \\
  1 & & & & & (1,0) & & (0,1) & & & & \\
  2 & & & & (2,0) & & {\color{grey} (1,1)} & & (0,2) & & & \\
  3 & & & (3,0) & & (2,1) & & (1,2) & & (0,3) & & \\
  4 & & (4,0) & & (3,1) & & {\color{grey} (2,2)} & & (1,3) & & (0,4) & \\
  5 & (5,0) & & (4,1) & & (3,2) & & (2,3) & & (1,4) & & (0,5)
  \end{array}
 \end{equation*}
 \caption{Total instanton number $n$ and net charge $q$ of $(m,\mb)$-instantons. The $\big(\tfrac{n}{2},\tfrac{n}{2}\big)$-con\-fig\-ur\-ations absent from the deformation in this section are greyed out.} \label{nqtable}
\end{table}

The structure of the solution for $\e^F$ up to $n=5$ is displayed in table~\ref{nqtable}. It is reminiscent of the resurgence triangles in e.g.\ \cite{DU}, but no resurgence occurs here since the asymptotic expansions \eqref{Kasympt} of the modified Bessel functions $K_\nu(z)$ are Borel-summable -- each instanton order is well-defined on its own.

The $n$-instanton functions $f_n$ are -- by construction -- interconnected of course, as is apparent from the appearance of the constants $A_k$ with $k<n$ in $f_n$. In fact, if we drop all homogeneous solutions to \eqref{Boxfn} for $n>1$ by setting $A_{n>1}=0$ so that only $A_1\equiv A$ remains, the higher instanton functions $f_n$ are determined completely in terms of the 1-instanton function $f_1$.

Let us study this case in the following. The $f_n$ in \eqref{fn} then take the form
 \begin{equation} \label{fng}
  f_n(x,w) = \frc{x}{2} \sum_{m+\mb=n}\+ \frc{A^m}{m!}\, \frc{\Ab^{\mb}}{\mb!}\ \e^{\+\I (m-\mb) w}\, g_{m,\mb}(x)
 \end{equation}
with $g_{m,\mb}(x)$ real functions symmetric in their indices and zero for $m=\mb$, such that
 \begin{align}
  f(x,w) = \frc{x}{2}\+ \Big[ & A\, \e^{\+\I w} g_{1,0}(x) + \frc{1}{2!}\, A^2\, \e^{\+2\I w} g_{2,0}(x) \notag \\[2pt]
  & + \frc{1}{3!}\+ \big( A^3\, \e^{\+3\I w} g_{3,0}(x) + 3\+ A^2 \Ab\, \e^{\+\I w} g_{2,1}(x) \big) \notag \\[2pt]
  & + \frc{1}{4!}\+ \big( A^4\, \e^{\+4\I w} g_{4,0}(x) + 4\+ A^3 \Ab\, \e^{\+2\I w} g_{3,1}(x) \big) \notag \\[2pt]
  & + \frc{1}{5!}\+ \big( A^5\, \e^{\+5\I w} g_{5,0}(x) + 5\+ A^4 \Ab\, \e^{\+3\I w} g_{4,1}(x) + 10\+ A^3 \Ab^2\, \e^{\+\I w} g_{3,2}(x) \big) \notag \\[2pt]
  & + \ldots \Big] + \cc\, .
 \end{align}
We have determined the $g_{m,\mb}$ exactly up to the 5-instanton level by solving equations \eqref{T1} to \eqref{T5} -- see appendix~\ref{app:gdgl}. The solutions read (dropping arguments for the sake of readability)
 \begin{align}
  n = 1:\quad  g_{1,0} & = K_1 \notag \\[.5ex]
  n = 2:\quad g_{2,0} & = 4\+ K_1\+ K_0 \notag \\[.5ex]
  n = 3:\quad g_{3,0} & = 9\+ K_1 \big( K_1^2 + 3\+ K_0^2 \big) \notag \\[.5ex]
  g_{2,1} & = K_1 \big( K_1^2 - K_0^2 \big) \notag \\[.5ex]
  n = 4:\quad g_{4,0} & = 256\+ K_1 \big( K_1^2\+ K_0^{} + K_0^3 \big) + 16\+ K_1^3 \big( K_2 - K_0 \big) \notag \\[.5ex]
  g_{3,1} & = 16\+ K_1 \big( K_1^2\+ K_0^{} - K_0^3 \big) + 4\+ K_1^3 \big( K_2 - K_0 \big) \notag \\[.5ex]
  n = 5:\quad g_{5,0} & = 625\+ K_1 \big( K_1^4 + 10\+ K_1^2\+ K_0^2 + 5\+ K_0^4 \big) + 50\+ K_1^3 \big( K_2 - K_0 \big) \big( K_2 + 14\+ K_0 \big) \notag \\[.5ex]
  g_{4,1} & = 81\+ K_1 \big( K_1^4 + 2\+ K_1^2\+ K_0^2 - 3\+ K_0^4 \big) + 18\+ K_1^3 \big( K_2 - K_0 \big) \big( K_2 + 6\+ K_0 \big) \notag \\[.5ex]
  g_{3,2} & = K_1 \big( K_1^4 - 2\+ K_1^2\+ K_0^2 + K_0^4 \big) + 2\+ K_1^3 \big( K_2 - K_0 \big) \big( K_2 + 2\+ K_0 \big)\ . \label{gupto5}
 \end{align}
Here we have used the identity \eqref{K2} to avoid explicit powers of $x$.

We make the following observations about the $g_{m,\mb}$ listed above:
 \newcounter{obs}
 \begin{list}{\arabic{obs}.}{\usecounter{obs} \parsep0ex \topsep1ex}
 \item All coefficients are integers.
 \item The sum of all coefficients in $g_{n,0}$ equals $(2n)^{n-1}$.
 \item The sum of all coefficients in $g_{m,\mb}$ with $m,\mb>0$ equals zero.
\end{list}
Since the leading term in the large $z$-expansion of $K_\nu(z)$ given in \eqref{Kasympt} is universal for all $\nu$, the second observation suggests that for large $x$
 \begin{equation} \label{gnasympt}
  g_{n,0}(x) \sim \e^{-nx}\+ (\pi/2x)^{n/2}\+ \big[ (2n)^{n-1} + O(x^{-1}) \big]\ .
 \end{equation}
We prove this asymptotic behaviour for all $n$ in appendix \ref{app:proof}. Furthermore, in the same appendix we show that for $m,\mb>0$
 \begin{equation} \label{gmasympt}
  g_{m,\mb}(x) \sim \e^{-nx}\+ (\pi/2x)^{n/2}\, x^{-1}\+ \big[\+ 2^{n-3}\+ (m-\mb)^2\, m^{m-2}\, \mb^{\mb-2} + O(x^{-1}) \big]\ .
 \end{equation}
This explains the third observation above since the factor $x^{-1}$ arises from the next-to-leading order terms in the expansions of the Bessel functions -- the leading order terms thus sum up to zero. It is remarkable that even though the expansion coefficients $a_1(\nu)$ in \eqref{Kasympt} that enter here are rational, we again only find integral prefactors to leading order. 
The reader is encouraged to verify \eqref{gmasympt} for the functions in \eqref{gupto5} using \eqref{Kasympt}.

Our results for $g_{m,\mb}$ listed in \eqref{gupto5} display a structure that is captured by the following empirical formula, valid at least up to $n=5$:\footnote{We have verified it also for $n=6$ but refrain from giving the lengthy results.}
 \begin{align}
  g_{m,\mb}(x) & = \frc{(-1)^{\mb}}{2}\, (m-\mb)^{n-1} \big[ (K_0 + K_1)^m\+ (K_0 - K_1)^{\mb} - (K_0 + K_1)^{\mb}\+ (K_0 - K_1)^m \big] \notag \\*[3pt]
  & \tab + h_{m,\mb}(x) \notag \\[2pt]
  & = \frc{(-1)^{\mb}}{2}\, (m-\mb)^{n-1} \sum_{k=0}^m \sum_{\ell=0}^{\mb} \binom{m}{k} \binom{\mb}{\ell} \big( (-1)^\ell - (-1)^k \big)\+ K_1^{k+\ell}\, K_0^{n-k-\ell} \notag \\*[-3pt]
  & \tab + h_{m,\mb}(x)\ , \label{gexact}
 \end{align}
where $h_{m,\mb}$ contains subleading terms of order $O(x^{-1} K_\nu^n)$ that are present for $n>3$:
 \begin{equation} \label{h}
  h_{m,\mb}(x) = 2^{n-3}\+ (m-\mb)^2\, K_1^3\, \sum_{k=1}^{n-3}\, k!\, (x^{-1} K_1)^k\, c_{m,\mb}^{(k)}(x)\ .
 \end{equation}
The functions $c_{m,\mb}^{(k)}$ are sums of monomials in $K_1$ and $K_0$ of order $n-3-k$. We have not found a complete expression for all $n$, but determine $c_{m,\mb}^{(n-3)}$ and $c_{m,\mb}^{(n-4)}$ in appendix~\ref{app:computeh}:
 \begin{equation} \label{c}
  c_{m,\mb}^{(n-3)}(x) = 1\ ,\quad c_{m,\mb}^{(n-4)}(x) = \big( m\+ (m-2) + \mb\+ (\mb-2) \big) K_0\ .
 \end{equation}
In the same appendix we also derive a partial result for $c_{m,\mb}^{(1)}$.

The formula \eqref{gexact} for $g_{m,\mb}$ is symmetric under interchange of $m$ and $\mb$ and vanishes for $m=\mb$ as required. For $\mb=0$ it reduces to
 \begin{align}
  g_{n,0}(x) & = \frc{n^{n-1}}{2}\+ \big[ (K_0 + K_1)^n - (K_0 - K_1)^n \big] + h_{n,0}(x) \notag \\[2pt]
  & = n^{n-1}\!\! \sum_{\text{odd}\,k\+\leq\+n}\!\! \binom{n}{k}\+ K_1^k\, K_0^{n-k} + h_{n,0}(x)\ .
 \end{align}
This immediately yields \eqref{gnasympt}, since to leading order the terms in square brackets produce the factor $2^n - 0^n$.
\bigskip

According to the asymptotic behaviour of the $g_{m,\mb}$ found above it is the pure instanton and anti-instanton configurations on the edges of the triangle in the $(m,\mb)$-table that dominate for each $n$, and the leading term in the large $x$-expansion of the functions $f_n(x,w)$ is captured by the expression in \eqref{gnasympt}. Moreover, due to the factor $(m-\mb)^2$ in \eqref{gmasympt}, at each level $n$ the contributions from $(m,\mb)$-instantons become ever smaller the more one moves into the interior of the triangle.

We can calculate the sum of \emph{all} leading instanton contributions. In terms of the original variables $r$ and $\vu$ and a rescaled integration constant $\alpha=-\sqrt{\smash[b]{\pi\,}}A$, we find
 \begin{equation} \label{leadsum}
  f(r,\vu\+) = \sum_{n=1}^\infty\+ f_n(r,\vu\+)\, \sim\, -\sqr\+ \re \sum_{n=1}^\infty\, \frc{(-n)^{n-1}}{n!}\, \big( \alpha\, r^{-1/4}\, \e^{-S}\+ \big)^n\, \big( 1 + O(r^{-1/2}) \big)
 \end{equation}
with complex 1-instanton action
 \begin{equation}
  S = 2\+ \sqr\% - \I\+ \vQ \cdot\% \vu\ .
 \end{equation}
Denoting the argument of the power series by $z=\e^{\+2\pi\I\+\tau}$, where
 \begin{equation}
  2\pi \re\tau = \vQ \cdot\% \vu + \arg\+ \alpha\ ,\quad 2\pi \im\tau = 2\+ \sqr\% + \ln\% \big( |\alpha|^{-1}\, r^{1/4} \big)\ ,
 \end{equation}
we are led to consider the expression
 \begin{equation} \label{W}
  W(z) = \sum_{n=1}^\infty\, \frc{(-n)^{n-1}}{n!}\, z^n\ .
 \end{equation}
This is the series expansion of the principal branch of the Lambert $W$-function \cite{NIST,CGHJK}\footnote{By coincidence I learned about the Lambert $W$-function from lectures on the Riemann zeta function by Andr\'e LeClair \cite{LC} while I was deriving \eqref{W}.}, which is defined as the solution to the algebraic equation $W\e^W=z$. Its radius of convergence is given by $\e^{-1}$, as is easily confirmed using the ratio test. Provided the constant $\alpha$ is not exponentially large, $z$ will reside in the disk of convergence for large $r$.

We conclude that
 \begin{equation}
  f(r,\vu\+) \sim -\sqr\+ \re W(\e^{\+2\pi\I\+\tau})\, \big( 1 + O(r^{-1/2}) \big)\ .
 \end{equation}
Here the factor $1+O(r^{-1/2})$ is to be understood as denoting our neglect of all subleading terms in each instanton sector. While within its disk of convergence the power series \eqref{W} is not just a formal one but an actual function, its real part taking values in the range $(-1\+,\+0.2785)$, $W(z)$ should be regarded as the generating function for the leading $n$-instanton contribution to $\e^{\+F}$.

\section{Conclusions} 

We have derived deformations of the classical moduli space metric of the universal hypermultiplet in Type II string compactifications which exhibit the form of membrane instanton corrections, by means of a suitable perturbation theory for the SU($\infty$) Toda equation. Our results significantly extend earlier such deformations in the literature by going beyond the one-instanton sector.

Since all membrane instanton corrections to the hypermultiplet moduli space are known in principle from a twistor construction \cite{A2}, it would be nice to compare our results with the latter. This should decide which of our deformations -- if any, as we do not claim to have found all possible ones corresponding to membrane instantons -- describes the universal hypermultiplet and fix the integration constants.

A feature of the deformation with two isometries constructed in section~\ref{sec:def2isom} which is surprising from a purely mathematical point of view (at least to the author) is the ubiquitous appearance of integers where naively one would expect rational numbers. This suggests a counting problem. In stringy instanton calculations, integers arise from counting BPS states, and it is encouraging that our formulae seem to reflect this, but there must exist an explanation independent of physics -- similar to the existence of integral Gopakumar--Vafa invariants for Calabi--Yau threefolds which are related to rational Gromov--Witten invariants (see e.g.\ \cite{Hetal}). The fact that the function $T(z)=-W(-z)$ is the exponential generating function for the number of labelled rooted trees \cite{CGHJK} leads us to speculate about a relation or combinatorial equivalence to problems in graph theory. This is supported by the appearance of the expression $m^{m-2}$ in the asymptotics of $g_{m,\mb}$, which according to Cayley's formula counts the number of trees on $m$ labelled vertices. A better understanding of the combinatorial properties of our perturbation theory might help in completing the deformations presented here into exact solutions to the Toda equation.

Ultimately, one would of course like to investigate the physical effects of quantum corrections to the hypermultiplet moduli space. First steps in this direction were taken in \cite{BM,DSTV}, where it was shown that membrane instantons allow for de~Sitter vacua in gauged $N=2$ supergravity, and more recently a proposal was put forward in \cite{K2} to derive natural inflation from instanton corrections to the universal hypermultiplet. We hope that our new results can contribute to these endeavours.
\vspace{3ex}

\textbf{Acknowledgements} \\[1ex] 
This work was begun a long time ago at the Institute for Theoretical Physics at Friedrich-Schiller-Uni\-ver\-si\-t\"at Jena and continued during visits to the II.\ Institute for Theoretical Physics at Universit\"at Hamburg. I would like to express my gratitude to Andreas Wipf and Jan Louis for their support and hospitality during these times. I also thank Sergei Alexandrov for insightful remarks on the first version of the preprint.

\appendix 

\section{Differential equations for $g_{m,\mb}(x)$} 
\label{app:gdgl}

In order to determine the functions $g_{m,\mb}(x)$ in the Ansatz \eqref{fng} for $f_n(x,w)$, the equations \eqref{Boxfn} need to be further decomposed with respect to the finer grading provided by the instanton/anti-instanton numbers $(m,\mb)$. Having made the $w$-dependence explicit in \eqref{fng}, the partial differential equations \eqref{Boxfn} then reduce to ordinary ones for $g_{m,\mb}$. Using the notation
 \begin{equation} \label{Boxq}
  \Box_q = x^2\+ \p_x^2 - x\+ \p_x - q^2\+ x^2\ ,
 \end{equation}
where $q=m-\mb$ as a result of applying $\Delta$ to $\e^{\+\I(m-\mb)w}$, the equations determining $g_{m,\mb}$ up to $n=5$ read
 \begin{align}
  n = 1:\quad  \Box_1\+ (x\, g_{1,0}) & = 0 \notag \\[.5ex]
  n = 2:\quad \Box_2\+ (x\, g_{2,0}) & = 8\+ x^2\+ g_{1,0}^2 \notag \\[.5ex]
  n = 3:\quad \Box_3\+ (x\, g_{3,0}) & = 18\+ \big( 4\+ x\, g_{1,0}^3 + 3\+ x^2\+ g_{1,0}\, g_{2,0} \big) \notag \\[.5ex]
  \Box_1\+ (x\, g_{2,1}) & = 2\+ \big( 4\+ x\, g_{1,0}^3 + x^2\+ g_{1,0}\, g_{2,0} \big) \notag \\[.5ex]
  n = 4:\quad \Box_4\+ (x\, g_{4,0}) & = 32\+ \big( 24\, g_{1,0}^4 + 24\+ x\, g_{1,0}^2\, g_{2,0}^{} + 4\+ x^2\+ g_{1,0}\, g_{3,0} + 3\+ x^2\+ g_{2,0}^2 \big) \notag \\[.5ex]
  \Box_2\+ (x\, g_{3,1}) & = 8\+ \big( 24\, g_{1,0}^4 + 12\+ x\, g_{1,0}^2\, g_{2,0}^{} + x^2\+ g_{1,0}\, g_{3,0} + 3\+ x^2\+ g_{1,0}\, g_{2,1} \big) \notag \\[.5ex]
  n = 5:\quad \Box_5\+ (x\, g_{5,0}) & = 50\+ \big( 192\+ x^{-1}\+ g_{1,0}^5 + 240\, g_{1,0}^3\, g_{2,0}^{} + 40\+ x\, g_{1,0}^2\, g_{3,0}^{} + 60\+ x\, g_{1,0}^{}\, g_{2,0}^2 \notag \\*
  & \mspace{48mu} + 10\+ x^2\+ g_{2,0}\, g_{3,0} + 5\+ x^2\+ g_{1,0}\, g_{4,0} \big) \notag \\[.5ex]
  \Box_3\+ (x\, g_{4,1}) & = 18\+ \big( 192\+ x^{-1}\+ g_{1,0}^5 + 144\, g_{1,0}^3\, g_{2,0}^{} + 24\+ x\, g_{1,0}^2\, g_{2,1}^{} + 16\+ x\, g_{1,0}^2\, g_{3,0}^{} \notag \\*
  & \mspace{48mu} + 12\+ x\, g_{1,0}^{}\, g_{2,0}^2 + 6\+ x^2\+ g_{2,0}\, g_{2,1} + 4\+ x^2\+ g_{1,0}\, g_{3,1} + x^2\+ g_{1,0}\, g_{4,0} \big) \notag \\[.5ex]
  \Box_1\+ (x\, g_{3,2}) & = 2\+ \big( 192\+ x^{-1}\+ g_{1,0}^5 + 96\, g_{1,0}^3\, g_{2,0}^{} + 36\+ x\, g_{1,0}^2\, g_{2,1}^{} + 4\+ x\, g_{1,0}^2\, g_{3,0}^{} \notag \\*
  & \mspace{40mu} + 12\+ x\, g_{1,0}^{}\, g_{2,0}^2 + 3\+ x^2\+ g_{2,0}\, g_{2,1} + x^2\+ g_{2,0}\, g_{3,0} + 2\+ x^2\+ g_{1,0}\, g_{3,1} \big)\ .
 \end{align}
Here we have used the symmetry relation $g_{m,\mb}=g_{\mb,m}$ in the equations with $\mb>0$. The solutions to the above inhomogeneous equations are given in \eqref{gupto5}, as can be easily verified using a computer algebra system.

\section{Asymptotic behaviour of $g_{m,\mb}(x)$} 
\label{app:proof}

We start with equation \eqref{gnasympt}, which we prove by induction. Let us assume therefore that the relation holds up to instanton order $n-1$. The functions in \eqref{gupto5} show that this assumption is true for the first few $n$. We then want to determine the leading term in $g_{n,0}(x)$, for which we have to solve the differential equation \eqref{Boxfn} for large $x$. The dominant terms on the right-hand side are contained in those multiplying the highest power of $r=x^2\%/4$. The latter arise from those partitions of $n$ with multiplicities $k_\ell$ for which $\sum_{\ell=1}^{n-1} k_\ell=2$, i.e.\ decompositions into two summands. There are only two kinds of possibilities -- either two different $k_\ell$ equal one or one $k_\ell$ equals two, with all other $k_\ell=0$. The latter can only happen for $\ell=n/2$. Thus, to leading order
 \begin{equation} \label{fnlead}
  \Box\+ f_n = -2\+ \Delta\+ \sum_{k=1}^{n-1}\+ f_k\, f_{n-k} + \dots\ .
 \end{equation}
Let us write the expansion \eqref{fng} in an obvious notation as
 \begin{equation}
  f_n =\! \sum_{m+\mb=n}\! f_{m,\mb}
 \end{equation}
with $f_{m,\mb}\propto\e^{\+\I(m-\mb)w}$. Inserting this into \eqref{fnlead} and using the induction hypothesis for $g_{k,0}$ with $k<n$, we then find
 \begin{align}
  \Box\+ f_{n,0} & = -2\+ \Delta\+ \sum_{k=1}^{n-1}\+ f_{k,0}\, f_{n-k,0} + \dots \notag \\
  & = -\frc{x^2}{2}\, \Delta\+ \sum_{k=1}^{n-1}\, \frc{A^k}{k!}\, \e^{\+\I k w}\+ g_{k,0}\ \frc{A^{n-k}}{(n-k)!}\, \e^{\+\I (n-k) w}\+ g_{n-k,0} + \dots \notag \\
  & = A^n\, \frc{x^2}{2}\, n^2\, \e^{\+\I n w}\, \sum_{k=1}^{n-1}\, \frc{1}{k!\, (n-k)!}\ g_{k,0}\, g_{n-k,0} + \dots \notag \\
  & \sim A^n\, \frc{x^2}{2}\, \e^{-nx}\+ (\pi/2x)^{n/2}\, \e^{\+\I n w}\, n^2\, \sum_{k=1}^{n-1}\, \frc{(2k)^{k-1}\, (2(n-k))^{n-k-1}}{k!\, (n-k)!} + \dots \notag \\
  & = \frc{A^n}{n!}\, \frc{x^2}{2}\, \e^{-nx}\+ (\pi/2x)^{n/2}\, \e^{\+\I n w}\, C_{n,0} + \dots \label{induct}
 \end{align}
with
 \begin{equation}
  C_{n,0} = 2^{n-2}\, n^2\, \sum_{k=1}^{n-1} \binom{n}{k}\, k^{k-1}\, (n-k)^{n-k-1} = (2n)^{n-1}\, n(n - 1)\ .
 \end{equation}
The last equality follows from evaluating the identity
 \begin{equation}
  \p_z^{n-2}\% f(z)^n = \frc{1}{2\+ (n-1)}\, \sum_{k=1}^{n-1} \binom{n}{k}\, \p_z^{k-1}\% f(z)^k\ \p_z^{n-k-1}\% f(z)^{n-k}
 \end{equation}
for $f(z)=\e^{\+z}$ at $z=0$.\footnote{For $f(z)=\e^{\+z/n}$ the same summands appear in recurrence relations relevant to the study of trees and random graphs \cite{CGHJK}.} We now consider the Ansatz
 \begin{equation}
  f_{n,0}(x,w) \sim \frc{A^n}{n!}\, \frc{x^p}{2}\, \e^{-nx}\+ (\pi/2x)^{n/2}\, \e^{\+\I n w}\+ N_{n,0} + \dots
 \end{equation}
with power $p$ and constant $N_{n,0}$ to be determined. When applying $\Box$ to $f_{n,0}$, we find
 \begin{equation} \label{Boxze}
  \Box_q\+ \big( \e^{-nx}\+ x^{p-n/2}\+ \big) = \e^{-nx}\+ x^{2+p-n/2}\+ \big[ (n^2 - q^2) + n\+ (n + 1 - 2p)\+ x^{-1} + O(x^{-2}) \big]
 \end{equation}
with $q=n$, where $\Box_q$ was defined in \eqref{Boxq}. The leading terms on both sides of \eqref{induct} then match if $p=1$ (this gives the overall factor of $x/2$ in \eqref{fng}) and $N_{n,0}=C_{n,0}/n(n-1)= (2n)^{n-1}$, which completes the proof of \eqref{gnasympt}.
\bigskip

We now turn to \eqref{gmasympt}, which we also prove by induction. The mild induction hypothesis is that up to order $n-1$ the $g_{m,\mb}$ with $n-1>m,\mb>0$ are subleading to $g_{n-1,0}$ for large $x$. Again, the functions in \eqref{gupto5} show that this assumption is true for the first few $n$. We then want to determine the leading asymptotic behaviour of $g_{m,\mb}$ with $\mb=n-m>0$. The dominant contribution to $\Box f_n$ is again given by \eqref{fnlead}. Due to the induction hypothesis the leading term in $f_k\,f_{n-k}$ of order $(m,\mb)$ is given by $f_{m,0}\,f_{0,\mb}$, which occurs twice -- all other terms involve the lower-order $f_{m,\mb}$ with $n-1>m,\mb>0$ which are subleading by assumption.\footnote{For example, $\Box f_{4,1} = -4\+\Delta \big( f_{4,0}\,f_{0,1} + f_{3,1}\,f_{1,0} + f_{2,1}\,f_{2,0} + \dots\big)$, where under the induction hypothesis the first term on the right-hand side contains the dominant contribution.} So we need to solve
 \begin{align}
  \Box\+ f_{m,\mb} & = -4\+ \Delta\+ \big( f_{m,0}\, f_{0,\mb} + \dots \big) = - x^2\, \frc{A^m}{m!}\, \frc{\Ab^{\mb}}{\mb!}\ \Delta\+ \big( \e^{\+\I m w}\+ g_{m,0}\ \e^{-\I \mb w}\+ g_{\mb,0} \big) + \dots \notag \\[.5ex]
  & \sim \frc{A^m}{m!}\, \frc{\Ab^{\mb}}{\mb!}\ \frc{x^2}{2}\, \e^{-nx}\+ (\pi/2x)^{n/2}\, \e^{\+\I (m-\mb)w}\, C_{m,\mb} + \dots \label{Boxfmmb}
 \end{align}
with
 \begin{equation}
  C_{m,\mb} = 2^{n-1}\+ (m-\mb)^2\, m^{m-1}\, \mb^{\mb-1}\ ,
 \end{equation}
where we have used the asymptotics of $g_{m,0}$ found above. We now make the Ansatz
 \begin{equation}
  f_{m,\mb}(x,w) \sim \frc{A^m}{m!}\, \frc{\Ab^{\mb}}{\mb!}\ \frc{x^p}{2}\, \e^{-nx}\+ (\pi/2x)^{n/2}\, \e^{\+\I (m-\mb)w}\+ N_{m,\mb} + \dots
 \end{equation}
with power $p$ and constant $N_{m,\mb}$ to be determined. When inserted into \eqref{Boxfmmb}, we obtain \eqref{Boxze} with $q=m-\mb$. Therefore, this time the leading term is the first on the right-hand side of \eqref{Boxze} and we achieve a match in \eqref{Boxfmmb} for $p=0$ and $N_{m,\mb}=C_{m,\mb}/ 4\+m\+\mb$. This proves the induction hypothesis and yields \eqref{gmasympt}.

\section{Subleading terms in $g_{m,\mb}(x)$} 
\label{app:computeh}

In this section we determine some properties of the functions $h_{m,\mb}$ in \eqref{gexact}. In particular, we compute the terms in $h_{m,\mb}$ -- and thus in  $g_{m,\mb}$ -- containing the two lowest explicit powers of $x$, i.e.\ the last summands in \eqref{h}, which are of the form $x^{3-n}K_\nu^n$ for $n>3$ and $x^{4-n}K_\nu^n$ for $n>4$. They arise from the two terms of lowest power of $r=x^2\%/4$ on the right-hand side of \eqref{Boxfn}. These are universal for all $n$ and are given by a single partition of $n$ each:
 \begin{equation} \label{Boxfnsub}
  \Box\+ f_n = -\frc{4}{n\, r^{n-2}}\ \Delta \big( f_1^n + n\+ r\+ f_1^{n-2} f_2^{} + \dots \big)\ ,
 \end{equation}
where the first term appears for $n>1$ and the second for $n>2$. While $f_n$ contains explicit powers of $r^{-1/2}$ for $n>3$, the powers of $r$ multiplying the $f_{n>3}$ on the right side of the above equation grow so quickly that the latter do not contribute to the terms we want to compute.

We are interested in the $(m,\mb)$-part of \eqref{Boxfnsub}. Using the expansion \eqref{fng}, the results for $f_1$ and $f_2$ following from \eqref{gupto5} and the identities
 \begin{align}
  (A + \Ab)^n & = n!\! \sum_{m+\mb=n} \frc{A^m}{m!}\, \frc{\Ab^{\mb}}{\mb!} \\[2pt]
  (A + \Ab)^{n-2}\+ (A^2 + \Ab^2) & = (n-2)!\! \sum_{m+\mb=n}\! (m^2 + \mb^2 - n)\, \frc{A^m}{m!}\, \frc{\Ab^{\mb}}{\mb!}\ ,
 \end{align}
we obtain with the notation \eqref{Boxq} and $q=m-\mb$
 \begin{align}
  \Box_q\+ (x\, g_{m,\mb}) & = 2^{n-1}\+ (m-\mb)^2\, \big[ (n-1)!\, x^{4-n}\+ K_1^n \notag \\*[2pt]
  & \tab + (n-2)!\, (m^2 + \mb^2 - n)\, x^{5-n}\+ K_1^{n-1} K_0^{} + \dots \big]\ , \label{Boxqg}
 \end{align}
again with the understanding that the two terms displayed on the right appear for $n>1$ and $n>2$, respectively. Consider now the function
 \begin{equation} \label{ansatzh}
  h_{m,\mb}(x) = 2^{n-3}\+ (m-\mb)^2\, K_1^3\+ \big[ a_{m,\mb}\, (x^{-1} K_1)^{n-3} + b_{m,\mb}\, (x^{-1} K_1)^{n-4} K_0 + \dots \big]\ .
 \end{equation}
Application of $\Box_q\,x$ produces a multitude of terms, those containing the two lowest explicit powers of $x$ being given by
 \begin{align}
  \Box_q\+ (x\, h_{m,\mb}) & = 2^{n-1}\+ (m-\mb)^2\, \big[ (n-1)(n-2)\, a_{m,\mb}\, x^{4-n}\+ K_1^n \notag \\*[2pt]
  & \tab + (n-2)\+ \big( n\, a_{m,\mb} + (n-3)\, b_{m,\mb} \big)\, x^{5-n}\+ K_1^{n-1} K_0^{} + \dots \big]\ .
 \end{align}
We achieve a match with \eqref{Boxqg} for
 \begin{equation}
  a_{m,\mb} = (n-3)!\ ,\quad b_{m,\mb} = (m^2 + \mb^2 - 2n)\+ (n-4)!\ ,
 \end{equation}
provided that $n\geq 3$ for the first term and $n\geq 4$ for the second. In the cases $n=3$ and $n=4$ the respective term in $h_{m,\mb}$ contains no explicit power of $x$ and is captured by $g_{m,\mb}-h_{m,\mb}$ in \eqref{gexact}. For $n>3$ and $n>4$, respectively, we read off from the above equations the functions $c_{m,\mb}^{(n-3)}$ and $c_{m,\mb}^{(n-4)}$ given in \eqref{c}.
\bigskip

The leading term in $h_{m,\mb}$ of the form $x^{-1}K_\nu^n$ is given by $c_{m,\mb}^{(1)}$. We have not found a way to determine it for all $n$, but we can at least derive the overall numerical coefficient for $m,\mb>0$. Let us denote the latter by $|c_{m,\mb}^{(1)}|$, such that\footnote{For instance, $c_{5,1}^{(1)}= \tfrac{1}{2}\big(27\+K_1^2+95\+K_0^2\big)$ and thus $|c_{5,1}^{(1)}|= 61=5^3-4^3$.} for large $z$
 \begin{equation}
  h_{m,\mb}(x) \sim 2^{n-3}\+ (m-\mb)^2\, \e^{-nx}\+ (\pi/2x)^{n/2}\, x^{-1}\+ \big(\+ |c_{m,\mb}^{(1)}| + O(x^{-1}) \big)\ .
 \end{equation}
Compare then the expansion of $g_{m,\mb}$ in \eqref{gexact} for $m,\mb>0$,
 \begin{align}
  & \e^{\+nx}\+ (\pi/2x)^{-n/2}\, g_{m,\mb}(x) \sim \frc{1}{2}\, (m-\mb)^{n-1} \Big[ \big( 2 + O(x^{-1}) \big)^m\+ \big( (2x)^{-1} + O(x^{-2}) \big)^{\mb} \notag \\*
  & \mspace{315mu} - (-1)^n\+ \big( 2 + O(x^{-1}) \big)^{\mb}\+ \big( (2x)^{-1} + O(x^{-2}) \big)^m \Big] \notag \\*[2pt]
  & \mspace{190mu} + 2^{n-3}\+ (m-\mb)^2\, x^{-1}\+ \big(\+ |c_{m,\mb}^{(1)}| + O(x^{-1}) \big) = \notag \\[6pt]
  & = x^{-1} \Big[ (m-\mb)^{n-1} \big( 2^{m-2}\, \delta_{\mb,1} - (-1)^n\, 2^{\mb-2}\, \delta_{m,1} \big) + 2^{n-3}\+ (m-\mb)^2\, |c_{m,\mb}^{(1)}| \Big]\! + O(x^{-2}) \notag \\[2pt]
  & = 2^{n-3}\+ (m-\mb)^2\, x^{-1} \Big[ (m-1)^{m-2}\, \delta_{\mb,1} + (\mb-1)^{\mb-2}\, \delta_{m,1} + |c_{m,\mb}^{(1)}| \Big]\! + O(x^{-2})\ ,
 \end{align}
with \eqref{gmasympt}. We conclude that
 \begin{equation} \label{c1}
  |c_{m,\mb}^{(1)}| = m^{m-2}\, \mb^{\mb-2} - (m-1)^{m-2}\, \delta_{\mb,1} - (\mb-1)^{\mb-2}\, \delta_{m,1}
 \end{equation}
for $m,\mb>0$. The latter restriction is necessary since we do not know the next-to-leading order terms in $g_{n,0}$ for arbitrary $n$. For $n=4$ and $n=5$, \eqref{c1} agrees with $|c_{m,\mb}^{(n-3)}|$ and $|c_{m,\mb}^{(n-4)}|$, respectively, as given in \eqref{c}. 
In these two cases, the latter also provide us with $|c_{n,0}^{(1)}|$.


\end{document}